\documentclass[12pt]{article}

\usepackage{graphics}
\usepackage{epsfig}
\input epsf

\title{Associated production of prompt photons and heavy quarks in off-shell gluon-gluon fusion}

\author{S.P.~Baranov$^a$, A.V.~Lipatov$^b$, N.P.~Zotov$^b$}

\begin{document}

\maketitle

\begin{center}

{\it $^a$\,P.N.~Lebedev Physics Institute,\\ 
119991 Moscow, Russia\/}\\[3mm]

{\it $^b$\,D.V.~Skobeltsyn Institute of Nuclear Physics,\\ 
M.V. Lomonosov Moscow State University,
\\119991 Moscow, Russia\/}\\[3mm]

\end{center}

\vspace{0.5cm}

\begin{center}

{\bf Abstract }

\end{center}

In the framework of the $k_T$-factorization approach, 
we study the production of prompt photons associated with heavy (charm 
and beauty) quarks in hadron-hadron collisions at high energies.
Our consideration is based on the amplitude for the production of a 
single photon associated with a quark pair in the fusion of two off-shell 
gluons. The total and differential cross sections are
presented and the conservative error analysis is performed.
Two sets of unintegrated gluon distributions in the proton
have been used in numerical calculation: the one
obtained from Ciafaloni-Catani-Fiorani-Marchesini evolution equation
and the other from Kimber-Martin-Ryskin prescription. 
The theoretical results are compared with recent
experimental data taken by the CDF collaboration at the Fermilab Tevatron.
Our analysis extends to specific angular 
correlations between the produced prompt photons and muons originating from
semileptonic decays of the final charmed or beauty quarks.
We point out the importance of such observables, which can serve as 
a crucial test for the unintegrated gluon densities in a proton.
Finally, we extrapolate the theoretical predictions to the CERN LHC energies.

This paper is dedicated to the memory of P.F.~Ermolov, who died on May 14, 2008.

\vspace{1cm}

\noindent
PACS number(s): 12.38.Bx, 13.85.Qk

\vspace{0.5cm}

\section{Introduction} \indent 

The production of prompt (or direct) photons in hadron-hadron collisions at the
Tevatron is a subject of intense studies on both the theoretical and experimental 
sides~[1--9]. Usually the photons are 
called "prompt" if they are coupled to the interacting quarks.
The theoretical and experimental investigations of such processes have 
provided a direct probe of the hard subprocess dynamics, since the produced 
photons are largely insensitive to the effects of final-state hadronization.
At the leading order of perturbative Quantum Chromodynamics (pQCD), 
prompt photons can be produced via quark-gluon Compton scattering 
or quark-antiquark annihilation and so, the cross sections of
these processes are strongly sensitive to the parton (quark and gluon) 
content of a proton\footnote{Also, the observed photon may arise 
from the so called fragmentation processes~[10]. This contribution will be discussed below 
in Section~2.}. The perturbative QCD calculations~[11] in the
next-to-leading order (NLO) approximation agree with the 
recent high-$p_T$ measurements~[6] within uncertainties (see also 
discussions in~[12--16]).
However, there are still open questions.
It was found~[1--4] that the shape of the measured cross section as a 
function of photon transverse energy $E_T$ is poorly described by the NLO pQCD 
calculations: the observed $E_T$ distribution is steeper than the 
predictions. This shape difference leads to a significant 
disagreement in the ratio of the cross sections calculated at different 
center-of-mass energies $\sqrt s = 630$ GeV and $\sqrt s = 1800$ GeV 
as a function of scaling variable $x_T = 2 E_T^\gamma/\sqrt s$.
It was demonstrated~[2, 3] that the disagreement in 
the $x_T$ ratio is difficult to explain with the
conventional theoretical uncertainties coming from the scale 
dependence and different parametrizations of the parton distributions.
In the NLO QCD approximation, the observed 
discrepancy can be reduced~[13, 17] by attributting some 
additional intrinsic transverse 
momentum $k_T$ to the incoming partons, which is usually assumed to 
have a Gaussian-like distribution. The average value of this $k_T$ 
increases from 
$\langle k_T \rangle \sim 1$ GeV to more than 
$\langle k_T \rangle \sim 3$ GeV~[13, 16] as 
the $\sqrt s$ increases from UA6 to Tevatron 
energies\footnote{The importance of including the gluon 
emission through the resummation 
formalism was recognized and only recently this approach 
has been developed~[17--21] for inclusive prompt photon production.}.

From our point of view, a  more adequate solution was found~[22--25] 
in the framework of the $k_T$-factorization approach~[26, 27].
This approach is based on the familiar 
Balitsky-Fadin-Kuraev-Lipatov (BFKL)~[28] 
or Ciafaloni-Catani-Fiorani-Marchesini (CCFM)~[29] gluon evolution 
equations and takes into account
the large logarithmic terms proportional to $\ln 1/x$.
It is believed that such terms give a significant 
contribution to the heavy quark production 
cross section at the conditions of modern colliders.
This contrasts with the usual 
Dokshitzer-Gribov-Lipatov-Altarelli-Parizi 
(DGLAP)~[30] strategy where only the large 
logarithmic terms proportional to $\ln \mu^2$ are taken into account.
The basic dynamical quantity of the $k_T$-factorization approach is 
the unintegrated (i.e., ${\mathbf k}_T$-dependent) gluon distribution 
${\cal A}(x,{\mathbf k}_T^2,\mu^2)$ which determines the probability to find a 
gluon carrying the longitudinal momentum fraction $x$ and the 
transverse momentum ${\mathbf k}_T$ at the probing scale $\mu^2$.
Similar to DGLAP, to calculate the cross sections of any 
physical process the unintegrated gluon density 
${\cal A}(x,{\mathbf k}_T^2,\mu^2)$ 
has to be convoluted~[26, 27] with the relevant partonic cross section 
which has to be taken off mass shell (${\mathbf k}_T$-dependent). 
Since the gluons in the initial state are not on-shell and are characterized by virtual 
masses (proportional to their transverse momentum), it also assumes a modification 
of their polarization density matrix~[26, 27]. In particular, the polarization 
vector of a gluon is no longer purely transversal, but acquires an admixture of 
longitudinal and time-like components. Other important properties of the 
$k_T$-factorization formalism are the additional contribution to the cross 
sections due to the integration over the ${\mathbf k}_T^2$ region above $\mu^2$
and the broadening of the transverse momentum distributions due to extra 
transverse momentum of the colliding partons.

In this approach, the treatment 
of the $k_T$-enhancement suggests a modification of the simple $k_T$ 
smearing picture described above: the 
transverse momentum $k_T$ of the incoming partons
is generated in the course of non-collinear parton evolution under 
control of the corresponding evolution equation.
First calculations of the inclusive prompt photon production 
at the Tevatron within the $k_T$-factorization formalism 
have been performed in~[22--25]. The calculations~[22--24] were
based on the $q + g^* \to \gamma + q$ and $q + \bar q \to \gamma + g$
off-shell matrix elements\footnote{In the calculations~[22], 
the usual on-shell matrix elements were embedded in 
precise off-shell kinematics.}. 
A reasonable agreement was found~[24] between the theoretical
predictions and the available D$\oslash$ and CDF experimental 
data~[1--6] in both the central and forward pseudo-rapidity $\eta^\gamma$ regions. 
Perfect agreement was found also in the ratio of the cross 
sections calculated at $\sqrt s = 630$ GeV and $\sqrt s = 1800$ GeV.
However, an important component of the calculations~[22--24] is the 
unintegrated quark distribution in a proton which at present
is available in the framework of Kimber-Martin-Ryskin 
(KMR)~[31] approach only\footnote{Unintegrated quark density was also
considered recently in~[32].}.
In contrast with~[22--24], the central part of our previous consideration~[25] 
is the off-shell gluon-gluon fusion subprocess $g^* + g^* \to \gamma + q \bar q$. 
At the price of considering the $2 \to 3$ rather than $2 \to 2$ matrix 
elements, the problem of unknown unintegrated quark
distributions has been reduced to the problem of gluon distributions.
In this way, since the gluons are only responsible for the appearance of the sea but 
not valence quarks, the contribution from the valence quarks should be 
calculated separately. Having in mind that the valence quarks are only 
important at large $x$, where the traditional DGLAP evolution is accurate and 
reliable, this contribution has been calculated within the usual collinear 
scheme based on $2\to 2$ partonic subprocesses and on-shell parton densities. 
Thus, the way proposed in~[25] enables us with making comparisons 
between the different parton 
evolution schemes and parametrizations of parton densities, in 
contrast with previous calculations~[22, 24] where such a comparison was not 
possible. It is important that the predictions~[25] based on the
off-shell gluon-gluon fusion matrix element $g^* + g^* \to \gamma + q\bar q$
and the KMR gluon density agree with the previous 
results~[24] based on the $2 \to 2$ subprocesses. 
This can be regarded as an additional 
proof of the consistency of the proposed method.

In the present paper we will apply the formalism~[25] described
above to investigate the prompt photon and associated heavy
(charm and beauty) quark production at high energies. 
The experimental data on the $\gamma + c$ and  $\gamma + b$ cross sections
as a function of photon transverse momentum $p_T$ have
been reported recently~[9] by the CDF collaboration.
Also, there are available data~[7, 8] on the associated prompt photon and
muon production at the Tevatron, 
where the final state muon originates from the semileptonic decay of a charmed
or beauty quark. Both these measurements are sensitive to the physics
beyound the Standard Model (SM), for example the production of 
excited quarks or gauge-mediated supersymmetry breaking (GMSB) with 
neutralinos radiatively decaying to gravitinos~[33].
Therefore, it is necessary to have a realistic estimation of the
associated $\gamma + Q$ or $\gamma + \mu$ production 
cross sections within QCD.
An additional motivation for our investigations is
the fact that these processes provide a direct 
probe of the off-shell matrix elements $g^* + g^* \to \gamma + q \bar q$
since there is no contribution from valence quarks.
In order to investigate the underlying dynamics in more detail, we study
the angular correlations between the transverse momenta of the
prompt photon and the final muon. These quantities are sensitive to 
the production mechanism and, also, powerful tests for 
the non-collinear evolution~[34].

The outline of our paper is following. In Section~2 we 
recall shortly the basic formulas of the $k_T$-factorization approach with a brief 
review of calculation steps. In Section~3 we present the numerical results
of our calculations and a discussion. Section~4 contains our conclusions.

\section{Theoretical framework} 
\subsection{Kinematics} \indent 

First, we recall in brief some technical details
of our previous
paper~[25] needed below.
We start from the kinematics (see Fig.~1). 
Let $p^{(1)}$ and $p^{(2)}$ be the four-momenta of the incoming protons and 
$p$ the four-momentum of the produced photon.
The initial off-shell gluons have the four-momenta
$k_1$ and $k_2$ and the final quark $Q$ and antiquark $\bar Q$ have the 
four-momenta $p_1$ and $p_2$ and the mass $m_Q$, respectively.
In the $p \bar p$ center-of-mass frame we can write
$$
  p^{(1)} = {\sqrt s \over 2} (1,0,0,1),\quad p^{(2)} = {\sqrt s \over 2} (1,0,0,-1), \eqno(1)
$$

\noindent
where $\sqrt s$ is the total energy of the process 
under consideration and we neglect the masses of the incoming protons.
The initial gluon four-momenta in the high energy limit can be written as
$$
  k_1 = x_1 p^{(1)} + k_{1T},\quad k_2 = x_2 p^{(2)} + k_{2T}, \eqno(2)
$$

\noindent 
where $k_{1T}$ and $k_{2T}$ are the transverse four-momenta.
It is important that ${\mathbf k}_{1T}^2 = - k_{1T}^2 \neq 0$ and
${\mathbf k}_{2T}^2 = - k_{2T}^2 \neq 0$. From the conservation laws 
we obtain the following relations:
$$
  {\mathbf k}_{1T} + {\mathbf k}_{2T} = {\mathbf p}_{1T} + {\mathbf p}_{2T} + {\mathbf p}_{T},
$$
$$
  x_1 \sqrt s = m_{1T} e^{y_1} + m_{2T} e^{y_2} + |{\mathbf p}_T| e^y, \eqno(3)
$$
$$
  x_2 \sqrt s = m_{1T} e^{-y_1} + m_{2T} e^{-y_2} + |{\mathbf p}_T| e^{-y},
$$

\noindent 
where $y$ is the rapidity of produced photon, 
$p_{1T}$ and $p_{2T}$ are the transverse four-momenta of final quark and antiquark, 
$y_1$, $y_2$, $m_{1T}$ and $m_{2T}$ are their center-of-mass rapidities and 
transverse masses, i.e. $m_{iT}^2 = m_Q^2 + {\mathbf p}_{iT}^2$.

\subsection{Cross section for associated $\gamma + Q$ production} \indent 

In general, according to the $k_T$-factorization theorem, the 
photon-quark associated production cross section  
can be written as a convolution
$$
  \displaystyle \sigma (p + \bar p \to \gamma + Q + X) = \int {dx_1\over x_1} {\cal A}(x_1,{\mathbf k}_{1 T}^2,\mu^2) d{\mathbf k}_{1 T}^2 {d\phi_1\over 2\pi} \times \atop 
  \displaystyle \times \int {dx_2\over x_2} {\cal A}(x_2,{\mathbf k}_{2 T}^2,\mu^2) d{\mathbf k}_{2 T}^2 {d\phi_2\over 2\pi} d{\hat \sigma} (g^* + g^* \to \gamma + Q \bar Q), \eqno(4)
$$

\noindent 
where $\hat \sigma(g^* + g^* \to \gamma + Q \bar Q)$ is the partonic cross section, 
${\cal A}(x,{\mathbf k}_{T}^2,\mu^2)$ is the unintegrated gluon distribution in a proton 
and $\phi_1$ and $\phi_2$ are the azimuthal angles of the incoming gluons.
The multiparticle phase space $\Pi d^3 p_i / 2 E_i \delta^{(4)} (\sum p^{\rm in} - \sum p^{\rm out} )$
is parametrized in terms of transverse momenta, rapidities and azimuthal angles:
$$
  { d^3 p_i \over 2 E_i} = {\pi \over 2} \, d {\mathbf p}_{iT}^2 \, dy_i \, { d \phi_i \over 2 \pi}. \eqno(5)
$$

\noindent
Using the expressions~(4) and~(5) we obtain the master formula:
$$
  \displaystyle \sigma(p + \bar p \to \gamma + Q) = \int {1\over 256\pi^3 (x_1 x_2 s)^2} |\bar {\cal M}|^2(g^* + g^* \to \gamma + Q \bar Q) \times \atop 
  \displaystyle \times {\cal A}(x_1,{\mathbf k}_{1 T}^2,\mu^2) {\cal A}(x_2,{\mathbf k}_{2 T}^2,\mu^2) d{\mathbf k}_{1 T}^2 d{\mathbf k}_{2 T}^2 d{\mathbf p}_{1 T}^2 {\mathbf p}_{2 T}^2 dy dy_1 dy_2 {d\phi_1\over 2\pi} {d\phi_2\over 2\pi} {d\psi_1\over 2\pi} {d\psi_2\over 2\pi}, \eqno(6)
$$

\noindent
where $|\bar {\cal M}|^2(g^* + g^* \to \gamma + Q\bar Q)$ is the off-mass shell 
matrix element squared and averaged over the initial gluon 
polarizations and colors, $\psi_1$ and $\psi_2$ are the 
azimuthal angles of the final state quark and antiquark, respectively.
Concerning the amplitude $g^* + g^* \to \gamma + Q \bar Q$,
there are eight Feynman diagrams which describe this partonic
subprocess at the leading order in $\alpha_s$ and $\alpha$ (see Fig.~2). 
The analytic expression for the $|\bar {\cal M}|^2(g^* + g^* \to \gamma + Q\bar Q)$ 
has been derived in our previous paper~[25].
We only mention here that, in accord with the $k_T$-factorization
prescription~[26, 27], the off-shell gluon spin density matrix has been 
taken in the form
$$
  \sum \epsilon^\mu (k_i) \epsilon^{*\,\nu} (k_i) = { k_{iT}^\mu k_{iT}^\nu \over {\mathbf k}_{iT}^2}. \eqno(7)
$$

\noindent 
In all other respects our calculations follow the standard Feynman rules.
If we average the expression~(6) over $\phi_{1}$ and $\phi_{2}$ 
and take the limit ${\mathbf k}_{1 T}^2 \to 0$ and ${\mathbf k}_{2 T}^2 \to 0$,
then we recover the relevant expression in the usual 
collinear approximation.

The multidimensional integration in~(6) has been performed
by the means of Monte Carlo technique, using the routine 
\textsc{Vegas}~[35]. The full C$++$ code is available from the 
authors on request\footnote{lipatov@theory.sinp.msu.ru}.
This code is practically identical to that used in~[25], with exception
that now we apply it to calculate the cross section of prompt photon 
and heavy quark (or rather decay muon) associated production.

\subsection{Photon isolation and fragmentation contribution} \indent 

In order to reduce huge background
from the secondary photons produced by the decays of $\pi^0$ and $\eta$ 
mesons, the isolation criterion is introduced in the experimental analyses.
This criterion is the following. A photon is isolated if the 
amount of hadronic transverse energy $E_T^{\rm had}$ deposited inside
a cone with aperture $R$ centered around the photon direction in the 
pseudo-rapidity and azimuthal angle plane is smaller than
some value $E_T^{\rm max}$:
$$
  \displaystyle E_T^{\rm had} \le E_T^{\rm max},\atop
  \displaystyle (\eta^{\rm had} - \eta)^2 + (\phi^{\rm had} - \phi)^2 \le R^2. \eqno(8)
$$

\noindent 
The CDF collaboration takes $R = 0.4$ and 
$E_T^{\rm max}= 1$~GeV in the experiment~[8, 9] and  
$R = 0.7$ and $E_T^{\rm max}= 2$~GeV in the earlier experiment~[7].

It is important that there is an additional mechanism of 
photon production not described above. It is
the fragmentation of a partonic jet into a single photon 
carrying a large fraction $z$ of the jet energy~[8]. These processes are 
described in terms of quark-to-photon $D_{q\to\gamma}(z,\mu^2)$ and
gluon-to-photon $D_{g\to\gamma}(z,\mu^2)$ fragmentation functions.  
However, the isolation condition (8) not only reduces the background 
but also significantly reduces the fragmentation components. 
It was shown~[36] that after applying the isolation cut the 
contribution from the fragmentation subprocesses is strongly 
suppressed (this contribution amounts 
to about 10\% of the visible cross section). 
Therefore in further analysis we will 
not consider the fragmentation component.

\section{Numerical results}
\subsection{Theoretical uncertainties} \indent 

There are several parameters which determine the overall 
normalization factor of the cross section~(6): 
the unintegrated gluon distribution in a proton 
${\cal A}(x,{\mathbf k}_T^2,\mu^2)$,
the factorization and renormalization scales 
$\mu_F$ and $\mu_R$ and the heavy quark mass $m_Q$.

Concerning the unintegrated gluon densities in a proton, 
we have tried here two different sets of them. These 
sets are widely discussed in the literature 
(see, for example, review~[37] for more information). 
Here we only shortly discuss their characteristic properties.

One set has been obtained~[38] recently
from the numerical solution of the CCFM equation. 
Function ${\cal A}(x,{\mathbf k}_T^2,\mu^2)$ is determined
by a convolution of the non-perturbative starting
distribution ${\cal A}_0(x)$ and the CCFM evolution kernel
denoted by $\tilde {\cal A}(x,{\mathbf k}_T^2,\mu^2)$:
$$
  {\cal A}(x,{\mathbf k}_T^2,\mu^2) = \int {d x'\over x'} {\cal A}_0(x') \tilde {\cal A}\left({x\over x'},{\mathbf k}_T^2,\mu^2\right). \eqno(20)
$$

\noindent
In the perturbative evolution the gluon splitting function
$P_{gg}(z)$ including non-singular terms (as it was described in~[39])
is applied. The input parameters in ${\cal A}_0(x)$
were fitted to reproduce the proton structure functions $F_2(x,Q^2)$.
An acceptable fit to the measured $F_2$ values was obtained~[38] with
$\chi^2/ndf = 1.83$ using statistical and uncorrelated systematic
uncertainties (compare to $\chi^2/ndf \sim 1.5$ in the collinear approach
at NLO).

Another set (the so-called KMR distribution) 
is the one which was originally proposed in~[31]. The KMR approach 
is a formalism to construct unintegrated gluon distribution from the 
known conventional parton (quark and gluon) densities. 
It accounts for the angular-ordering (which comes from 
the coherence effects in gluon emission) as well as the main part of the 
collinear higher-order QCD corrections. The key observation here 
is that the $\mu$ dependence of the unintegrated parton distribution 
enters at the last step of the evolution
and therefore single scale evolution equations 
can be used up to this 
step\footnote{In the numerical calculations we have used the 
standard GRV~(LO) parametrizations~[40] of the collinear quark and gluon 
densities.}. 

Significant theoretical uncertainties are connected with the
choice of the factorization and renormalization scales. The first of them
is related to the evolution of the gluon distributions, the other is 
responsible for the strong coupling constant $\alpha_s(\mu^2_R)$.
As it is often done, we choose the 
renormalization and factorization scales to be equal: 
$\mu_R = \mu_F = \mu = \xi |{\mathbf p}_{T}|$.
In order to investigate the scale dependence of our 
results we will vary the scale parameter
$\xi$ between $1/2$ and~2 about the default value $\xi = 1$.

In the numerical calculations we set the charm and beauty quark masses 
to $m_c = 1.4$~GeV and $m_b = 4.75$~GeV. 
We have checked that the uncertainties which come 
from these quantities are negligible in comparison with the uncertainties
connected with the unintegrated gluon distributions.
For completeness, we use the LO formula for the strong 
coupling constant $\alpha_s(\mu^2)$ with $n_f = 4$ 
active quark flavors at 
$\Lambda_{\rm QCD} = 200$~MeV (so that $\alpha_s(M_Z^2) = 0.1232$).
Note that we use the special choice $\Lambda_{\rm QCD} = 130$~MeV 
in the case of CCFM gluon ($\alpha_s(M_Z^2) = 0.1187$), 
as it was originally proposed in~[38]. 

\subsection{Associated $\gamma + Q$ production at Tevatron} \indent

We are now in a position to present our numerical results.
Figs.~3 and~4 confront the $\gamma + c$ and
$\gamma + b$ production cross sections calculated as a function of 
the photon transverse energy $E_T$ with the preliminary experimental 
data~[9] taken by the CDF collaboration at the Tevatron Run~II ($\sqrt s = 1960$~GeV).
These data refer to the central kinematic region defined by
$|\eta^\gamma| < 1$.
The solid and dotted histograms correspond to the 
results obtained with the CCFM and KMR unintegrated gluon densities, 
respectively. The upper and lower dashed 
histograms correspond to the scale variations in CCFM density 
as it was described above.
We find that the CCFM-evolved unintegrated gluon density 
reproduces well the data within the theoretical and experimental
uncertainties, and that the KMR density tends to underestimate the data
in wide $E_T$ range.
A similar effect was observed also in the case of inclusive
prompt photon production~[25].
The difference between the CCFM and KMR predictions
is directly connected with the small-$x$ behaviour of
these gluon densities and demonstrates the importance
of leading $\ln 1/x$ terms.
Of course, the scale uncertainties of our predictions
are significant. The latter can be reduced
by considering the ratio of the cross sections $\gamma + c$ to 
$\gamma + b$. This ratio is a subject of special interest: 
one could expect from the ratio of the quark charges that
the $\gamma + c$ events should be 4 times more often
than the $\gamma + b$ events. In addition to that,
there must be extra suppression of the $\gamma + b$ events
due to heavier $b$ mass (in the $g + g \to \gamma + Q\bar Q$ approach)
or smaller beauty content in the proton sea (in the $g + Q \to \gamma + Q$ approach).
Our prediction for the ratio $\sigma (\gamma + c)/\sigma (\gamma + b)$ is
shown in Fig.~5 in comparison with the CDF data~[9].
Both the CCFM and KMR gluon densities predict this ratio 
to be equal to $6:1$ or $7:1$ in a wide $E_T$ range.
This result is consistent with the measurement~[9].

There are also available CDF data~[7, 8] on the muons which originate from the
semileptonic decays of charmed or beauty quarks. 
The experimental data~[8] refer to the kinematic region 
$p_T^\mu > 4$~GeV, $|\eta^\gamma| < 0.9$, $|\eta^\mu| < 1.0$
and $\sqrt s = 1800$~GeV. In Fig.~6 we show the $\gamma + \mu$
cross section as a function of photon transverse momentum $p_T$.
The contributions from both the $\gamma + c$ and $\gamma + b$ events 
have been taken into account.
To produce muons from charmed and beauty quarks in our 
theoretical calculations,
we first convert them into a $D$ or $B$ hadrons using
the Peterson fragmentation function~[41] and then 
simulate their semileptonic decay according to the
standard electroweak theory\footnote{Of course, the muon transverse momentum spectra are 
sensitive to the fragmentation functions. However, this dependence is 
expected to be small as compared with the uncertainties coming from the unintegrated 
gluon densities in a proton.}. As usual, we set the fragmentation 
parameter $\epsilon$ to $\epsilon_c = 0.06$ and $\epsilon_b = 0.006$.
The branching fractions $f(c \to \mu)$ and $f(b \to \mu)$ were set to 
$f(c \to \mu) = 0.09$ and $f(b \to \mu) = 0.1078$~[42].
One can see that in the case of $\gamma + \mu$ production, our 
predictions with the CCFM gluon density slightly overestimate
the data but still agree with them within the uncertainties.
The collinear NLO QCD calculations~[43] give similar description
of the data. The results obtained with the KMR density lie below the
measurements and are similar to those~[8] obtained from 
the \textsc{Pythia}~Monte Carlo~[44].

An important point of our consideration is the 
investigation of the angular correlations between the 
prompt photon and heavy quark.
It is well known that studying these correlations gives 
additional insight into the production dynamics and, in particular, 
into the effective contributions from higher-order QCD processes.
For example, the lowest-order QCD production diagrams $g + Q\to \gamma + Q$ 
contain only the photon $\gamma$ and heavy quark $Q$ in the final state.
Therefore, the distribution over $\Delta \phi = \phi_\gamma - \phi_Q$ 
must be simply a delta function $\delta(\Delta \phi - \pi)$ since 
the produced particles are back-to-back in the transverse plane and 
are balanced in $p_T$ due to momentum conservation.
When higher-order QCD processes are considered, the
presence of additional quarks and/or gluons in the final state
allows the $\Delta \phi$ distribution to be more spread 
and the heavy quark transverse momenta more asymmetric.
In the $k_T$-factorization formalism, taking into 
account the non-vanishing initial gluon
transverse momentum ${\mathbf k}_{T}$ leads to 
the violation of back-to-back kinematics even at leading order.
However, using the $2 \to 3$ matrix elements 
instead the $2 \to 2$ ones (as it was described above) 
makes the difference between the 
$k_T$-factorization predictions and the collinear approximation of QCD 
(in $\alpha_{em}\alpha_s^2$ approximation) not well pronounced.

The associated $\gamma + \mu$ cross section as a function 
of the azimuthal angle $\Delta \phi (\gamma - \mu)$ has been measured in~[7]. 
The data on the normalized differential cross 
section $(1/\sigma)\,d\sigma/d\Delta \phi (\gamma - \mu)$ have been presented.
These data refer to the kinematic region 
$17 < E_T < 40$~GeV, $p_T^\mu > 4$~GeV, $|\eta^\gamma| < 0.9$, $|\eta^\mu| < 1.0$
and $\sqrt s = 1800$~GeV. Our theoretical prediction compared to the data
are shown in Fig.~7. We find here a number of the interesting points.
First, the shapes of histograms 
predicted by the CCFM and KMR gluon densities are very different 
from each other. 
The CCFM gluon reproduces well the shape of the measured $\Delta \phi$ distribution,
although tends to slightly overestimate the data at $\Delta \phi \sim \pi$,
while the KMR gluon density is unable to describe the data anywhere.
The observed shape difference is in contrast with the transverse
momentum spectra, where both the unintegrated gluon
distributions under consideration demonstrate a (more or less) similar behaviour.
This fact clearly indicates that the $\gamma + \mu$ 
cross section as a function of 
$\Delta \phi$ is very sensitive to the details of the non-collinear evolution.
A similar observation has been made earlier~[34] in the case
of $b$-quark hadroproduction at the Tevatron.
Thus, futher theoretical and experimental studying such
quantities can give us the possibility to additional constrain the
unintegrated gluon densities. However, we should mention that the
behaviour of the $\Delta \phi$ distribution at $\Delta \phi \sim 0$ 
depends sensibly on the photon 
isolation criteria. In particular, it depends on the parameters 
$R$ and $E_T^{\rm max}$ determining the cone isolation~(8).
Our predictions for LHC energy are shown in Figs.~8 and~9.

In conclusion, we would like to stress a number
of important achievements shown by the $k_T$-factorization
approach. As a general feature, the model behaviour is
found to be perfectly compatible with the available
data on the heavy quark production as well as on the 
production of prompt photons and various quarkonium 
states at modern colliders~[45].
It is important that the $k_T$-factorization approach succeeds 
in describing the polarization phenomena observed in both $p\bar p$
and $ep$ interactions (see, for example,~[46] and references therein). 
The underlying physics is
essentially related to the initial gluon off-shellness,
which dominates the gluon polarization properties and
has a considerable impact on the kinematics. 
So, we believe that the $k_T$-factorization formalism holds a 
possible key to understanding the production dynamics 
at high energies. Finally, once again we
would like to point out the fundamental role of angular
correlations which can serve as an important and crucial test
discriminating the different non-collinear evolution schemes.

\section{Conclusions} \indent 

We have tried a theoretical approach proposed in~[25] to the 
associated production of prompt photons and heavy (charmed or beauty) quark 
in hadronic collisions at high energies. 
Our approach is based on the $k_t$-factorization scheme, which, 
unlike many early calculations~[13, 16], provides solid theoretical grounds 
for adequately taking into account the effects of initial parton 
momentum. The central part of our consideration is the off-shell gluon-gluon 
fusion subprocess $g^* + g^* \to \gamma + q\bar{q}$. 
At the price of considering the $2 \to 3$ rather than $2 \to 2$ 
matrix elements, we have reduced 
the problem of unknown unintegrated quark distributions to the 
problem of gluon distributions. 
This way enables us with making comparisons between the different parton 
evolution schemes and parametrizations of parton densities, in 
contrast with previous calculations~[22--24] where such a comparison was not 
possible (for the lack of unintegrated quark distributions except KMR). 

We have calculated the total and differential $\gamma + Q$ 
and $\gamma + \mu$ cross sections (where muon originates 
from the semileptonic decay of the heavy quark $Q$)
and have made comparisons to the recent CDF 
experimental data. In the numerical analysis we have used 
the unintegrated gluon densities obtained from the  
CCFM evolution equation and from the KMR prescription.
It was demonstrated that the CCFM-evolved gluon density reproduces
the Tevatron data very well, whereas the KMR gluon density is in disagreement with
them. We especially point out the fundamental role
of the angular correlations between the particles in the final state. These 
quantities can serve as a crucial test for 
the unintegrated gluon densities in a proton.
Finally, we extrapolate the theoretical predictions 
to LHC energies\footnote{When the present paper
was prepared for publication, the D$\oslash$ collaboration
presented~[47] the first measurement of the prompt 
photon and hadronic jet associated production. The triple
differential cross section $d\sigma/d p_T d\eta^\gamma d\eta^{\rm jet}$ 
has been determined in the different kinematic regions. 
It could also be very useful to analyze it
using the off-shell $g^* + g^* \to \gamma + q\bar{q}$ 
matrix elements. We are planning presentation
of this work in the forthcoming publications.}.

\section{Acknowledgements} \indent 

We thank H.~Jung for his encouraging interest, very helpful discussions
and for providing the CCFM code for 
unintegrated gluon distributions. 
The authors are very grateful to 
DESY Directorate for the support in the 
framework of Moscow --- DESY project on Monte-Carlo
implementation for HERA --- LHC.
A.V.L. was supported in part by the grants of the president of 
Russian Federation (MK-438.2008.2) and Helmholtz --- Russia
Joint Research Group.
Also this research was supported by the 
FASI of Russian Federation (grant NS-8122.2006.2)
and the RFBR fundation (grant 08-02-00896-a).

\newpage 

\begin{figure}
\begin{center}
\epsfig{figure=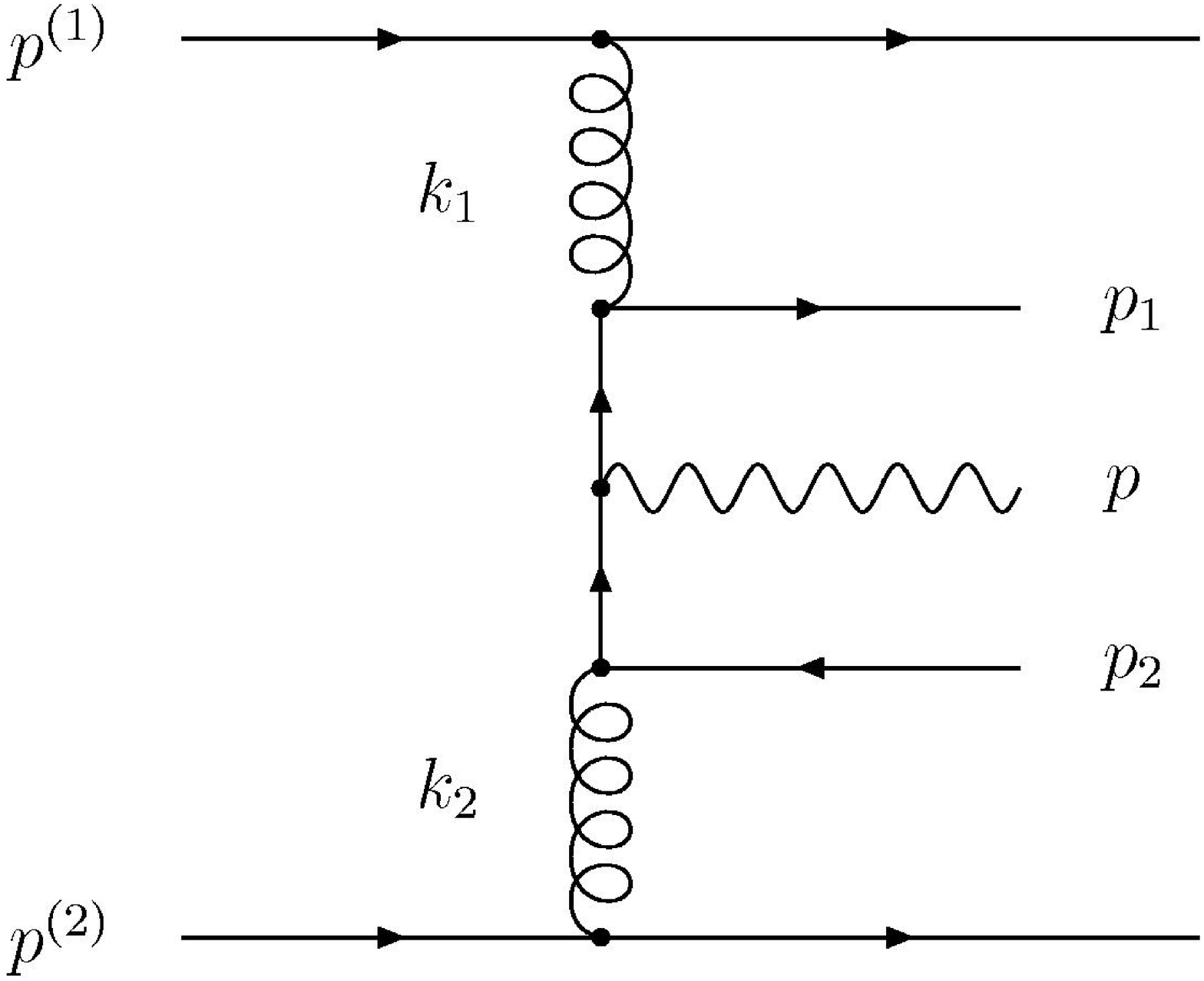, width = 8cm}
\caption{Kinematics of the $g^* + g^* \to \gamma + Q \bar Q$ process.}
\label{fig1}
\end{center}
\end{figure}

\newpage

\begin{figure}
\begin{center}
\epsfig{figure=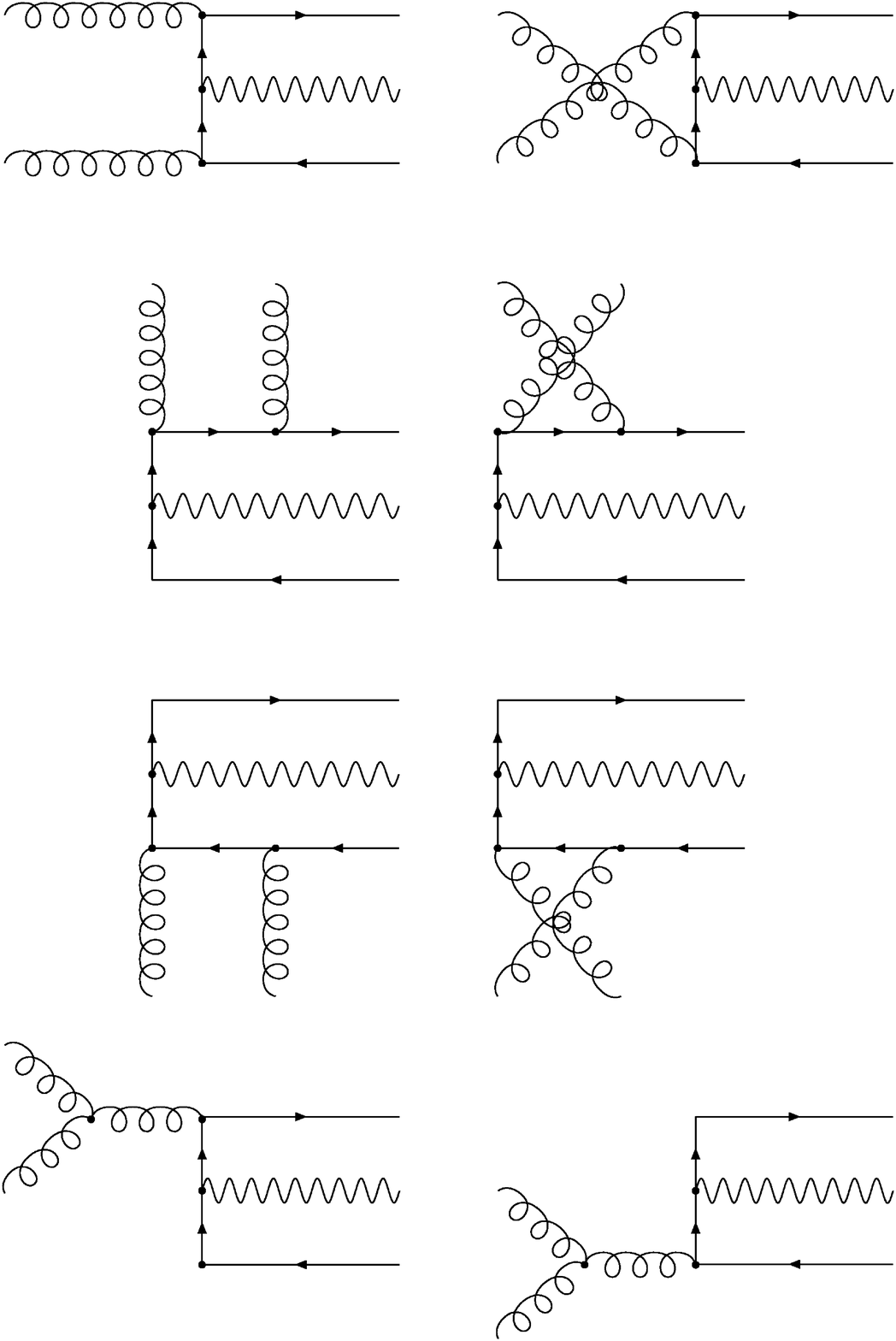, width = 14cm}
\caption{Feynman diagrams which describe the partonic
subprocess $g^* + g^* \to  \gamma + Q \bar Q$ at the leading order 
in $\alpha_s$ and $\alpha$.}
\label{fig2}
\end{center}
\end{figure}

\newpage

\begin{figure}
\begin{center}
\epsfig{figure=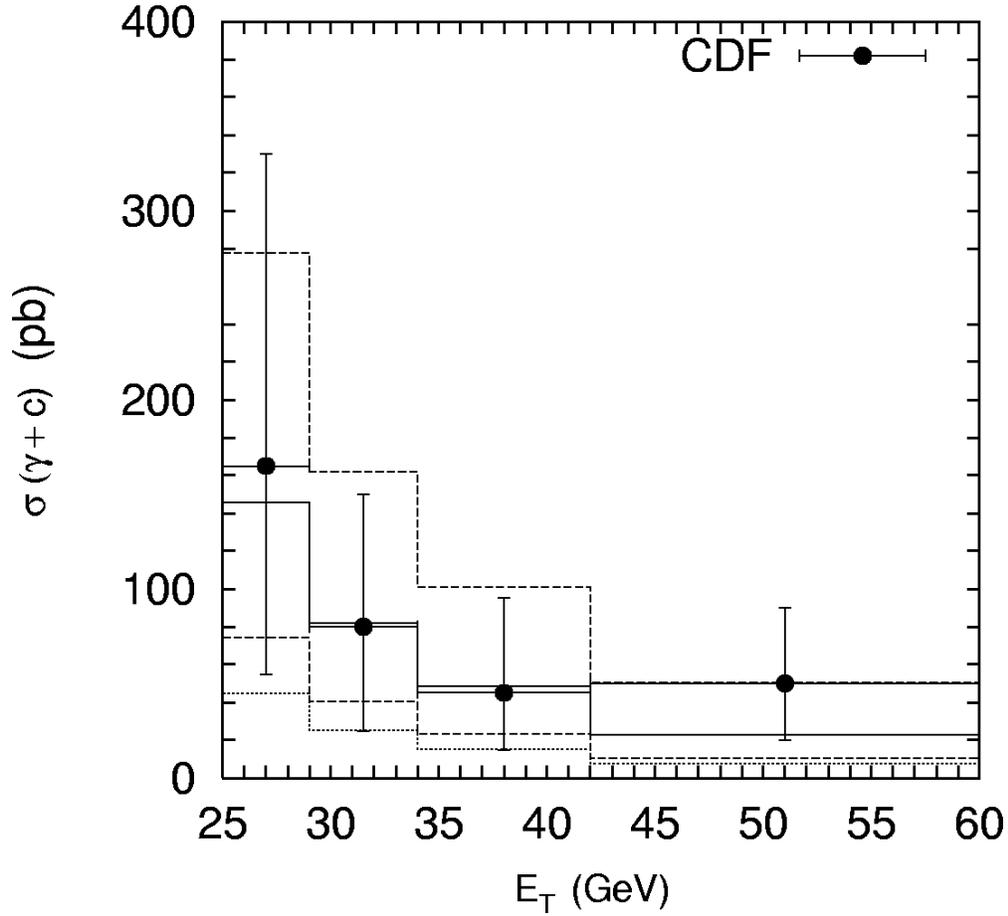, width = 18cm}
\caption{The $\sigma(\gamma + c)$ cross section
as a function of photon transverse energy $E_T$ at
$|\eta| < 1.0$ and $\sqrt s = 1960$~GeV. 
The solid and dotted histograms correspond to the CCFM 
and KMR gluon densities, respectively, with the default 
scale $\mu = E_T$. The upper and lower dashed histograms
correspond to the scale variation in the CCFM distribution. 
The experimental data are from CDF~[9].}
\end{center}
\label{fig3}
\end{figure}

\newpage

\begin{figure}
\begin{center}
\epsfig{figure=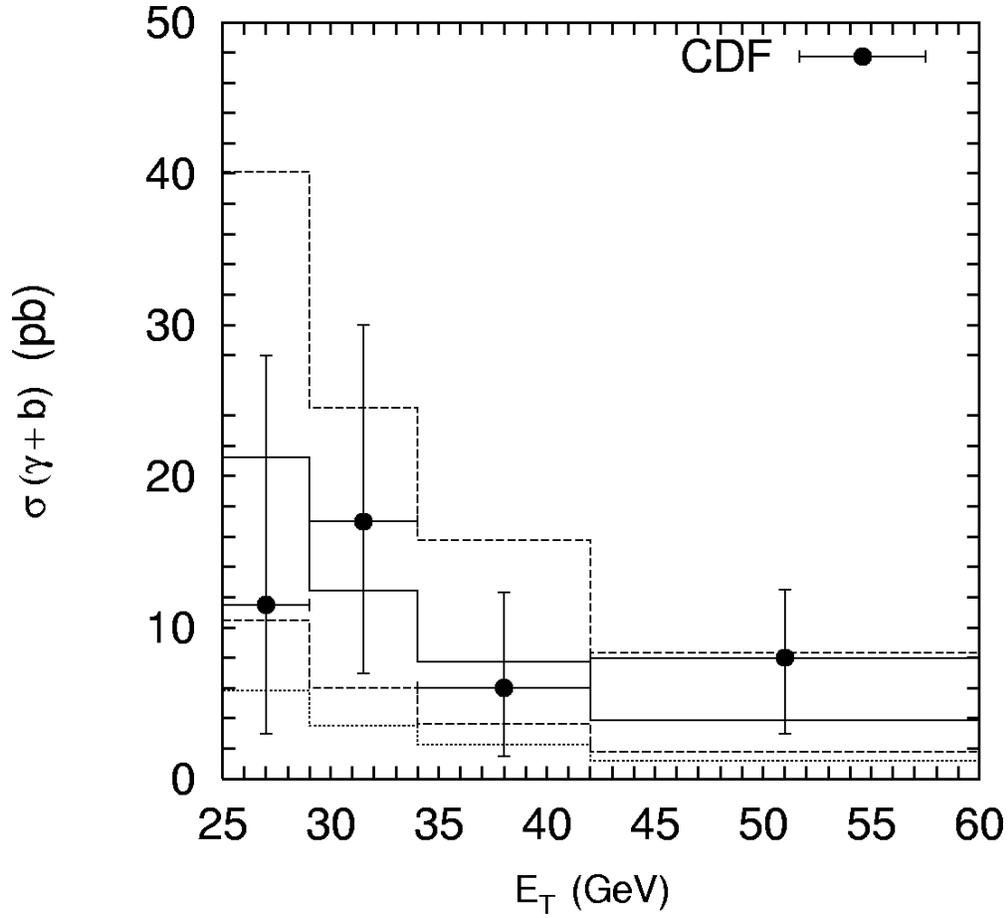, width = 18cm}
\caption{The $\sigma(\gamma + b)$ cross section
as a function of photon transverse energy $E_T$ at
$|\eta| < 1.0$ and $\sqrt s = 1960$~GeV. 
Notations of histograms are the same as in Fig.~3.
The experimental data are from CDF~[9].}
\end{center}
\label{fig4}
\end{figure}

\newpage

\begin{figure}
\begin{center}
\epsfig{figure=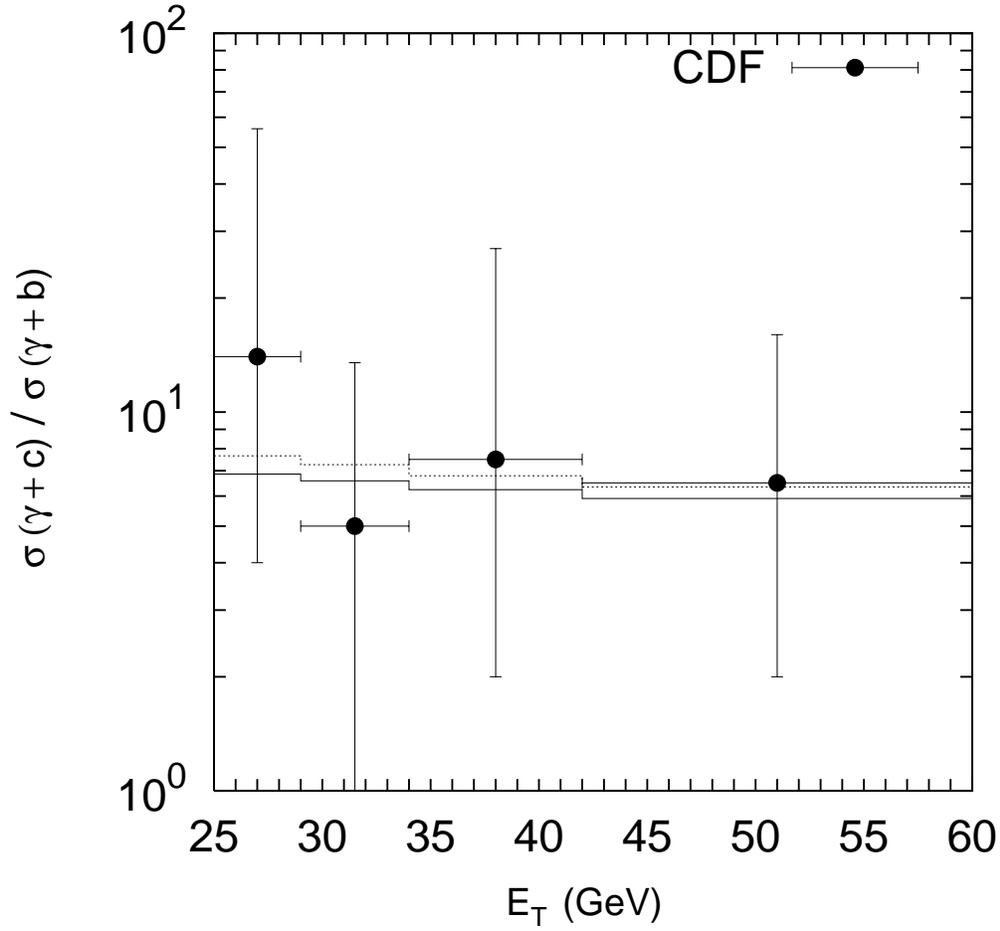, width = 18cm}
\caption{The ratio of $\gamma + c$ to $\gamma + b$ 
cross sections as a function of photon transverse energy $E_T$ at
$|\eta| < 1.0$ and $\sqrt s = 1960$~GeV. 
The solid and dotted histograms correspond to the CCFM 
and KMR gluon densities, respectively, with the default 
scale $\mu = E_T$.
The experimental data are from CDF~[9].}
\end{center}
\label{fig5}
\end{figure}

\newpage

\begin{figure}
\begin{center}
\epsfig{figure=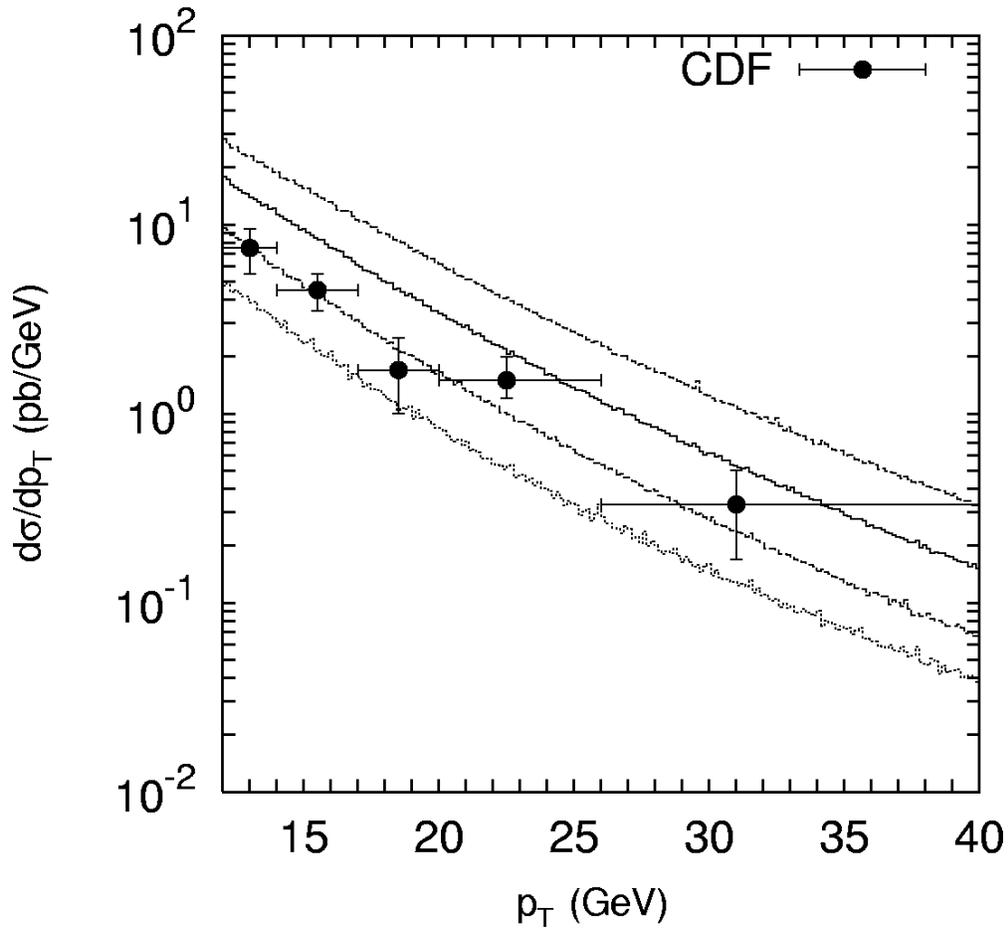, width = 18cm}
\caption{The differential cross section $d\sigma/d p_T$
for associated prompt photon and muon hadroproduction calculated at 
$p_T^\mu > 4$~GeV, $|\eta^\gamma| < 0.9$, $|\eta^\mu| < 1.0$ and
$\sqrt s = 1800$~GeV.
Notations of histograms are the same as in Fig.~3.
The experimental data are from CDF~[8].}
\end{center}
\label{fig6}
\end{figure}

\newpage

\begin{figure}
\begin{center}
\epsfig{figure=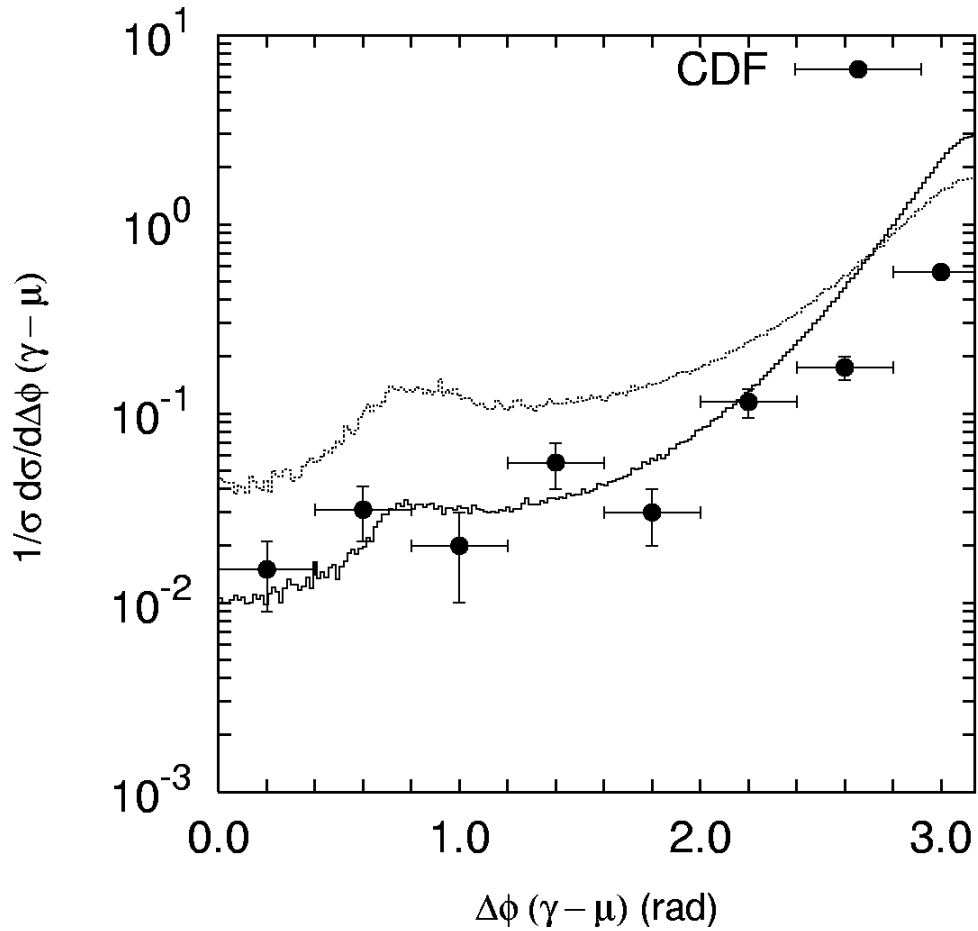, width = 18cm}
\caption{The difference in azimuthal angle $\Delta \phi (\gamma - \mu)$ 
between the photon and muon calculated at 
$17 < p_T < 40$~GeV, $p_T^\mu > 4$~GeV, $|\eta^\gamma| < 0.9$, $|\eta^\mu| < 1.0$ and
$\sqrt s = 1800$~GeV.
Notations of histograms are the same as in Fig.~5.
The experimental data are from CDF~[7].}
\end{center}
\label{fig7}
\end{figure}

\newpage

\begin{figure}
\begin{center}
\epsfig{figure=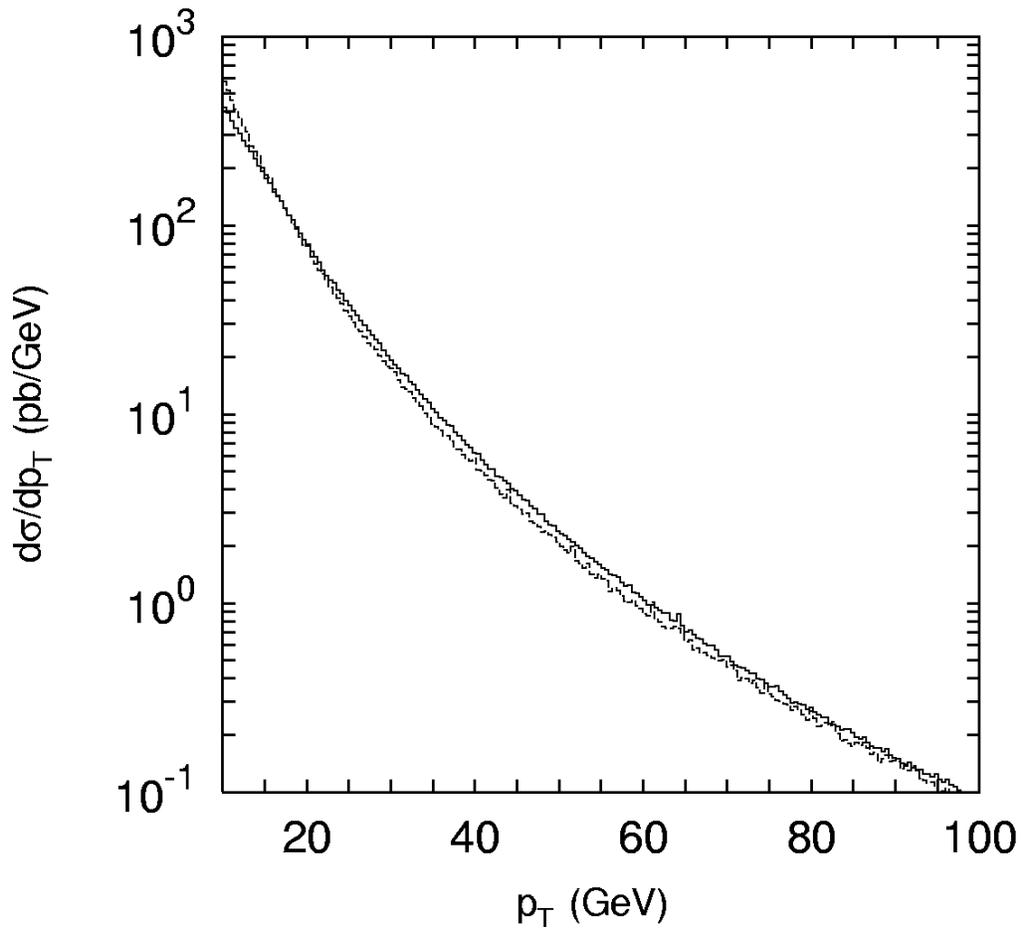, width = 18cm}
\caption{The differential cross section $d\sigma/d p_T$
for associated prompt photon and muon hadroproduction calculated at 
$p_T^\mu > 4$~GeV, $|\eta^\gamma| < 2.5$, $|\eta^\mu| < 2.5$ and
$\sqrt s = 14$~TeV.
Notations of histograms are the same as in Fig.~5.
The isolation criterion as in~[8, 9] was applied.}
\end{center}
\label{fig8}
\end{figure}

\newpage

\begin{figure}
\begin{center}
\epsfig{figure=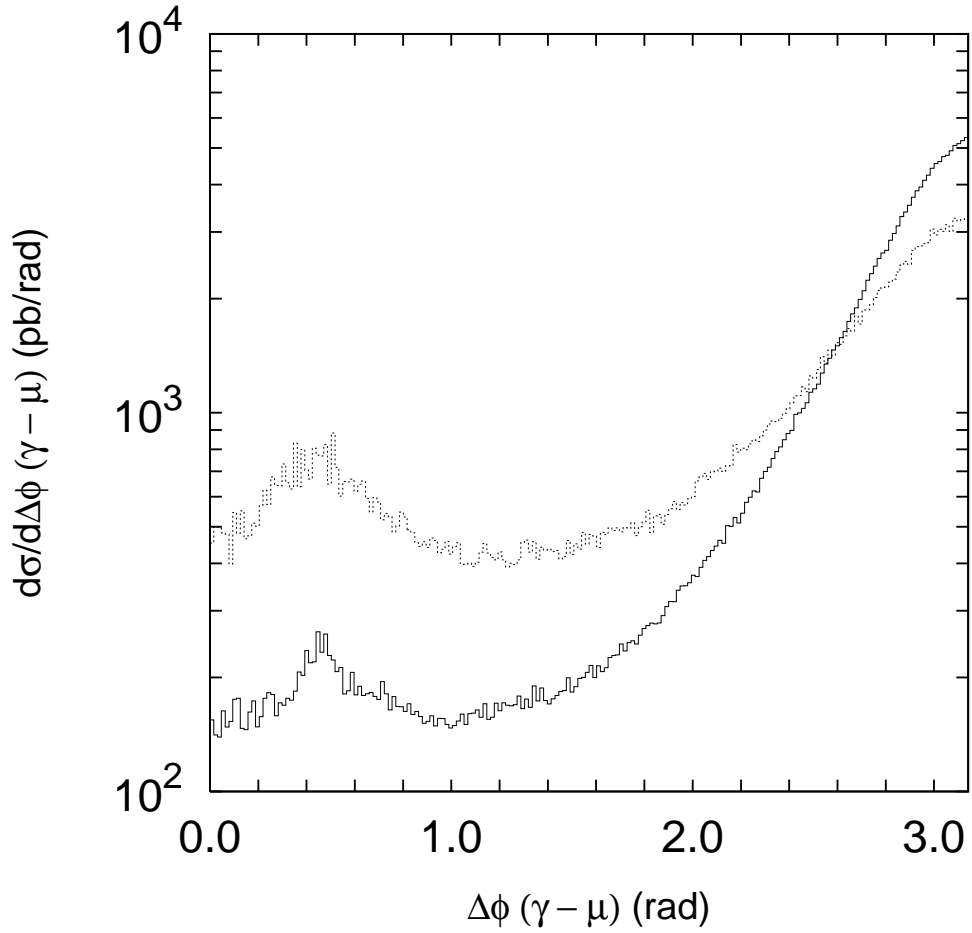, width = 18cm}
\caption{The difference in azimuthal angle $\Delta \phi (\gamma - \mu)$ 
between the photon and muon calculated at 
$10 < p_T < 100$~GeV, $p_T^\mu > 4$~GeV, $|\eta^\gamma| < 2.5$, $|\eta^\mu| < 2.5$ and
$\sqrt s = 14$~TeV. Notations of histograms are the same as in Fig.~5. 
The isolation criterion as in~[8, 9] was applied.}
\end{center}
\label{fig9}
\end{figure}

\end{document}